\documentclass[aps,prb,twocolumn,english,showpacs,superscriptaddress,amssymb,amsfonts]{revtex4}
\usepackage[T1]{fontenc}
\usepackage[latin9]{inputenc}
\usepackage{amsmath}
\usepackage{dsfont}
\usepackage{amstext}

\usepackage{tocvsec2}

\usepackage{subcaption}
\usepackage{amssymb}
\usepackage{amsbsy}
\usepackage{amsthm}
\usepackage{epsfig}
\usepackage{framed}
\usepackage{graphicx}
\usepackage{bbm}
\usepackage{hyperref}
\usepackage{color}
\usepackage{multirow}

\newcommand{\Ref}[1]{Ref.~\onlinecite{#1}}

\newcommand{\bst}{{\mathcal{T}}}

\newcommand{\ie}{{\emph{i.e.~}}}
\makeatletter

\newcommand{\Rmnum}[1]{\expandafter\@slowromancap\romannumeral #1@}
\makeatother
\newcommand{\imth}{\hspace{1pt}\mathrm{i}\hspace{1pt}}

\newcommand{\eg}{{\emph{e.g.~}}}

\newcommand{\mbz}{{\mathbb{Z}}}
\newcommand{\bea}{\begin{eqnarray}}
\newcommand{\eea}{\end{eqnarray}}
\newcommand{\bpm}{\begin{pmatrix}}
\newcommand{\epm}{\end{pmatrix}}
\newcommand{\bal}{\begin{aligned}}
\newcommand{\eal}{\end{aligned}}

\newcommand{\expval}[1]{\langle{#1}\rangle}

\makeatother

\usepackage{babel}
\begin{document}
\title{Magnon band topology in spin-orbital coupled magnets:\\ classification and application to $\alpha$-RuCl$_3$}

\author{Fuyan Lu}
\author{Yuan-Ming Lu}
\affiliation{Department of Physics, The Ohio State University, Columbus, OH 43210, USA}

\begin{abstract}
In spite of flourishing studies on the topology of spin waves, a generic framework to classify and compute magnon band topology in non-collinear magnets is still missing. In this work we provide such a theory framework, by mapping an arbitrary linear spin wave into a local free-fermion Hamiltonian with exactly the same spectrum, symmetry implementation and band topology, which allows for a full classification and calculation on any topological properties of magnon bands. We apply this fermionization approach to honeycomb Kitaev magnet $\alpha$-RuCl$_3$, and show the existence of topologically protected magnon band crossings, and field-induced magnon Chern bands under small magnetic fields.
\end{abstract}


\maketitle




\section{Introduction}

The discovery of topological insulators revealed a large class of electronic materials, which support symmetry protected surface states as a manifestation of nontrivial bulk topological properties\cite{Hasan2010,Qi2011,Chiu2016,Armitage2018}. The success of topological band theory in electronic systems leads to a natural question: can similar topological phenomena appear in the energy bands of quasiparticle excitations in a bosonic system? Indeed various topological bands and protected surface states have been engineered in mechanical\cite{Kane2014,Lubensky2015,Huber2016} and photonic systems\cite{Ozawa2018}. Meanwhile, the prevalent family of magnetically ordered materials provide another ideal platform, where the spin wave excitations can realize various topological bands and magnon surface surface states. Although lots of discoveries have been made recently\cite{Chisnell2015,Chisnell2016,Yao2017,Shindou2013,Shindou2013a,Zhang2013b,Lisenkov2014,Romhanyi2015,Lawler2016,Kim2016a,Owerre2016,Owerre2016a,Li2016,Attig2017,Laurell2017,Nakata2017,Li2017a,Li2017b,Iacocca2017,Joshi2017,Hwang2017,Bao2018,Joshi2018} on the topology of spin waves (or magnons), unlike in electronic systems, a systematic framework to understand magnon band topology is still lacking.

In this work, we resolve this issue by establishing an exact mapping between a (bosonic) linear spin wave theory and a (fermionic) free-electron system. This mapping preserves locality and all physical symmetries of the linear spin wave, so that the corresponding pair of linear spin wave (or magnon) and free fermion systems share exactly the same bulk spectrum and band topology. For example, the magnons of a generic spin-orbit coupled magnet are mapped to Bogoliubov quasiparticles in an electronic superconductor. In another more familiar example, magnons in a collinear magnet with $U(1)$ spin conservation are mapped to electrons in an insulator with $U(1)$ charge conservation.

This ``fermionization'' approach establishes a correspondence between linear spin waves and well-understood free-fermion Hamiltonians, thus allowing us to fully classify and compute band topology of spin wave (or magnon) excitations. To demonstrate its power, we apply this formulation to study magnon band topology in layered honeycomb ``Kitaev material'' $\alpha$-RuCl$_3$, where spin-orbit couplings play an important role\cite{Rau2016,Winter2017,Hermanns2018}. We show that the zigzag order in $\alpha$-RuCl$_3$ exhibits symmetry-protected magnon band touchings, and they can be lifted by an external magnetic field, giving rise to magnon Chern bands with chiral magnon edge states. We study the evolution of magnon bands in $\alpha$-RuCl$_3$ by varying the applied magnetic field, and obtain a phase diagram of magnon band topology as a function of the magnetic field.


\section{Fermionization and band topology of linear spin waves}

In an ordered magnet, spins precess around the direction of the local magnetic field, which is determined by the local moment via the magnetic interaction. The dynamics of a generic linear spin wave (LSW) is determined by the following equation of motion (e.o.m.):
\bea\label{spin wave eom:0}
-\imth\frac{\text{d}s_i^\alpha}{\text{d}t}=\sum_{j,\beta}\big(\sigma_y\cdot{\bf R}\big)_{i\alpha,j\beta}s_j^\beta,~~~\alpha,\beta=x,y
\eea
where $\vec s_i$ denotes the fluctuation of spin ${\bf S}_i$ on site $i$ from its ordered moment $\expval{{\bf S}_i}$
\bea
\vec s_i={\bf S}_i-\expval{{\bf S}_i}
\eea
For convenience we have chosen a local coordinate frame where the ordered moment $\expval{{\bf S}_i}$ on every site $i$ points to $\hat z$ direction, and Pauli matrices $\vec\sigma$ act on the $\alpha=x,y$ indices. As detailed in supplemental materials, the matrix ${\bf R}$ is fully determined by the original spin Hamiltonian and the ordered moments $\{\expval{{\bf S}_i}\}$ of the LSW.

Eq. (\ref{spin wave eom:0}) can be viewed as a Schrodinger equation, where eigenvalues of matrix $\sigma_y\cdot{\bf R}$ determine the magnon (or spin wave) spectrum. The fact that ${\bf R}={\bf R}^\ast$ is a real matrix implies a ``particle-hole symmetry'' of the eigenvalues of $\sigma_y{\bf R}$: a positive eigenvalue $\omega_j>0$ must appear in pair with a negative eigenvalue $-\omega_j<0$. In a generic spin-orbit coupled magnet, the ``Hamiltonian'' matrix $\sigma_y\cdot{\bf R}$ is \emph{not Hermitian}. This roots in the difference between bose and fermi statistics, as compared to a Hermitian Hamiltonian in any free-fermion system.

\begin{table*}[tb]
\centering
\begin{tabular} {|c||c|c||c|}
\hline
&Spin wave e.o.m.&Holstein-Primakoff approach&Free fermion systems\\
\hline
Physical problems&$-\imth\partial_t\vec s=(\sigma_y{\bf R})\cdot\vec s$&$\mathcal{H}_\text{H-P}=\phi^\dagger{\bf H}_b\phi$&$\mathcal{H}_\text{FF}=\psi^\dagger{\bf H}_f\psi$\\
\hline
Variables&$\vec s=(s_i^x,s_i^y)^T$&$\phi=(b_i,b_i^\dagger)^T$&$\psi=\big(\frac{f_i+f_i^\dagger}2,\imth\frac{f_i^\dagger-f_i}2\big)^T$\\
\hline
``Hamiltonian'' matrix&${\bf R}={\bf R}^\dagger={\bf R}^\ast$&${\bf H}_b=U^\dagger{\bf R}U$&${\bf H}_f=\sqrt{\bf R}\cdot\sigma_y\cdot\sqrt{\bf R}$\\
&&$U=e^{\imth\frac\pi4(1-\sigma_z)}e^{\imth\frac\pi4\sigma_y}$&\\
\hline
Diagonalization &$(\sigma_y{\bf R})\cdot({\bf B}e^{\imth\frac\pi4\sigma_x})=$&
$(\sigma_z{\bf H}_b)\cdot{\bf W}_b=$&${\bf H}_f\cdot{\bf W}_f=$\\
of Hamiltonian&$({\bf B}e^{\imth\frac\pi4\sigma_x})\cdot(\sigma_z\otimes\mathbf{\Omega})$&${\bf W}_b\cdot(\sigma_z\otimes\mathbf{\Omega})$&${\bf W}_f\cdot(-\sigma_z\otimes\mathbf{\Omega})$\\
\hline
``Particle-hole''&${\bf R}={\bf R}^\ast$,&${\bf H}_b=\sigma_x{\bf H}_b^\ast\sigma_x$,&${\bf H}_f^\ast=-{\bf H}_f$,\\
symmetry&${\bf B}={\bf B}^\ast$.&${\bf W}_b^\ast=\sigma_x{\bf W}_b\sigma_x$.&${\bf W}_f^\ast={\bf W}_f\sigma_x$.\\
\hline
Wavefunction&${\bf B}^\dagger\sigma_y{\bf B}=\sigma_y$&${\bf W}_b^\dagger\sigma_z{\bf W}_b=\sigma_z$&${\bf W}_f^\dagger{\bf W}_f=1$\\
normalization&&&\\
\hline
Relation between&${\bf B}_{2N\times2N}\in Sp(2N,\mathbb{R})$&${\bf W}_b=U^\dagger{\bf B}U$&${\bf W}_f=\sqrt{\bf R}{\bf B}U\mathbf{\Omega}^{-\frac12}$\\
wavefunctions&&&$=U\sqrt{{\bf H}_b}{\bf W}_b\mathbf{\Omega}^{-\frac12}$\\
\hline
Diagonal form&${\bf R}=\sigma_y{\bf B}\mathbf{\Omega}{\bf B}^\dagger\sigma_y$&${\bf H}_b=\sigma_z{\bf W}_b\mathbf{\Omega}{\bf W}_b^\dagger\sigma_z$&
${\bf H}_f={\bf W}_f(-\sigma_z\mathbf{\Omega}){\bf W}_f^\dagger$\\
\hline
Unitary symmetry $g\in G_s$&$[O_g,{\bf R}]=0,$&$[U^\dagger O_gU,{\bf H}_b]=0,$&$O_g{\bf H}_fO_g^\dagger={\bf H}_f$\\
$O_g\in Sp(2N,\mathbb{R})$&$[O_g,\sigma_y]=0$&$[U^\dagger O_gU,\sigma_z]=0$&\\
\hline
Anti-unitary symmetry $h\in G_s$&$[O_h,{\bf R}]=0,$&$[U^\dagger O_hU,{\bf H}_b]=0,$&$O_h{\bf H}_fO_h^\dagger=-{\bf H}_f={\bf H}_f^\ast$\\
$O_h\in SO(2N)$&$\{O_h,\sigma_y\}=0$&$\{U^\dagger O_hU,\sigma_z\}=0$&\\
\hline
\end{tabular}
\caption{The relation between semiclassical LSW equation of motion (e.o.m.), Holstein-Primakoff boson formalism of LSWs, and the corresponding free-fermion systems with the same spectrum. The ``particle-hole symmetry'' is a redundancy of the formulation rather than a physical symmetry. We have defined $\mathbf{\Omega}_{i,j}\equiv\delta_{ij}\omega_j\geq0$ as the diagonal matrix of non-negative magnon frequencies, and $U=e^{\imth\frac\pi4(1-\sigma_z)}e^{\imth\frac\pi4\sigma_y}$ is a unitary rotation in the $\vec\sigma$ space satisfying $U^\ast=U\sigma_x=\sigma_zU$. This fermionization map is well defined for gapped magnon spectra with $\det{\bf R}\neq0$.}
\label{tab:map}
\end{table*}

To map the LSW into a free-fermion system, the key step is the following similarity transformation:
\bea\label{fermionization map}
{\bf R}^{1/2}\cdot\big(\sigma_y{\bf R}\big)\cdot{\bf R}^{-1/2}=\sqrt{\bf R}~\sigma_y\sqrt{\bf R}={\bf H}_f
\eea
Note that for a generic spin-orbit coupled magnet, the gapped magnon spectrum indicates that matrix ${\bf R}$ is positive-definite and hence its square root is uniquely defined. This ``fermionization map'' generates a free fermion Hamiltonian ${\bf H}_f$ with exactly the same spectrum as the boson Hamiltonian. It's straightforward to check that ${\bf H}_f$ is Hermitian and particle-hole symmetric. Moreover it preserves the same symmetries as the LSW, as shown in TABLE \ref{tab:map}. While every gapped LSW system can be mapped to a short-ranged free-fermion model, not all free-fermion Hamiltonians have their LSW counterpart. As proved in Supplemental materials, the ground state for any free-fermion counterpart ${\bf H}_f$ of the LSW Hamiltonian ${\bf R}$ must be a trivial product state, since it can always be adiabatically connected to the fermion atomic insulator without closing the gap while preserving the same symmetries. This is consistent with topological triviality of a magnetically-ordered ground state, and rules out the possibility of any zero-energy magnon surface states protected by symmetries.

Although the magnetic ground state at zero energy is topologically trivial, each magnon band can still exhibit nontrivial topology and symmetry-protected surface states at finite energy. Using known K-theory classification\cite{Kitaev2009,Wen2012,Morimoto2013} for free fermions\cite{Chiu2016}, the fermionization map (\ref{fermionization map}) allows us to fully classify and compute magnon band topology with various global and crystalline symmetries, by looking into their free-fermion partners. The main results are summarized in TABLE \ref{tab:classification} for various remaining symmetries of the magnetic orders, together with possible materials to realize these topological magnons.

In addition to the topology for each gapped magnon band separated from other bands, one can also classify the topology of symmetry-protected band touchings in a LSW spectrum, using a dimensional reduction approach introduced in free-fermion systems\cite{Horava2005,Wang2012,Matsuura2013,Zhao2013,Chiu2016}. Specifically protected point nodes in $d$-dimensional magnets are classified by $(d-1)$-dimensional gapped magnon bands, while line nodes are classified by $(d-2)$-dimensional gapped magnon bands. In a simplest case most familiar in the literature, for collinear magnetic orders with a $U(1)$ spin rotational symmetry along local $\hat z$-axis, we have $[\sigma_y,{\bf R}]=0$ and hence the fermionized Hamiltonian is nothing but the LSW matrix ${\bf R}={\bf H}_f$. Here the LSW theory reduces to diagonalizing a free-fermion hopping Hamiltonian, which is the case for Cu(1,3-bdc)\cite{Chisnell2015,Chisnell2016} and Cu$_3$TeO$_6$\cite{Li2017b,Yao2017,Bao2018}. While many symmetries do give rise to topological magnon bands and band touchings, we found that collinear and coplanar magnetic orders always have topologically trivial magnons (and hence no magnon surface states), if the combination $\tilde\bst$ of time reversal and certain spin rotation is preserved in the magnetic order.

\begin{table*}[tb]
\centering
\begin{tabular} {|c|c||c||c|c|c|}
\hline
Physical&Magnetic orders&Classifying&$d=1$&$d=2$&$d=3$\\
symmetry&and realizations&Space&&&\\
\hline
No symmetry&Ferro(i)magnets (FM) w/ SOC&$\mathcal{C}_d$&0&$\mbz$&0\\
&Cu(1,3-bdc)\cite{Chisnell2015,Chisnell2016}&&&&point\cite{Li2016}\\
\hline
$U(1)_{S^z}$&Chiral collinear FM w/o SOC&$\mathcal{C}_d$&0&$\mbz$&0\\
&&&&&point\\
\hline
$U(1)_{S^z}\rtimes Z_2^{\tilde\bst}$&Non-chiral collinear FM w/o SOC&$\mathcal{R}_{8-d}$&0&0&0\\
$\tilde\bst=e^{\imth\pi S_y}\cdot\bst$&&&&&\\
\hline
$Z_2^{\tilde\bst}$&Coplanar orders w/o SOC&$\mathcal{R}_{8-d}$&0&0&0\\
$\tilde\bst=e^{\imth\pi S_y}\cdot\bst$&&&&&\\
\hline
2-fold rotation $C_2$&&$(\mathcal{C}_{d})^2$&0&$\mbz\times\mbz$&0\\
&&&&point&point/line\\
\hline
Magnetic rotation&&$\mathcal{R}_{4-d}$&0&$\mbz_2$&$\mbz_2$\\
$\tilde C_2=C_2\cdot\bst$&&&&point&point/line\\
\hline
Mirror $R_a$&Zigzag order in $\alpha$-RuCl$_3$&$\mathcal{C}_{d+1}$&$\mbz$&0&$\mbz$\\
&Red dots in FIG. \ref{fig:symmetry}&&&point&line\\
\hline
Magnetic mirror&Zigzag order in $\alpha$-RuCl$_3$&$\mathcal{R}_{2-d}$&$\mbz_2$&$\mbz$&0\\
${\tilde R_a}\equiv\bst\cdot R_a$&Green dots in FIG. \ref{fig:symmetry}&&&point&point/line\\
\hline
Inversion $I$&&&$\mbz$&$\mbz\times\mbz$&$\mbz$\\
&&&&point&point/line\\
\hline
Magnetic inversion&Cu$_3$TeO$_6$\cite{Li2017b,Yao2017,Bao2018}&$\mathcal{R}_{d}$&$\mbz_2$&$\mbz_2$&0\\
${\tilde I}\equiv\bst\cdot I$&&&&point&point/line\\
\hline
Magnetic translation&Neel antiferromagnet w/ SOC&&0&0&$\mbz_2$\\
${\tilde T_1}\equiv\bst\cdot T_1$&Yellow dots in FIG. \ref{fig:symmetry}&&point&line&sheet\\
\hline
\end{tabular}
\caption{Classification of magnon band topology in various magnetic orders protected by the unbroken symmetries, obtained using K-theory\cite{Kitaev2009,Teo2010,Wen2012,Morimoto2013}. For every symmetry class, the first row shows the classification of each gapped magnon band, while the 2nd row shows the possible types of symmetry-protected magnon band touchings.}
\label{tab:classification}
\end{table*}

\begin{figure}
  \includegraphics[width=0.9\linewidth]{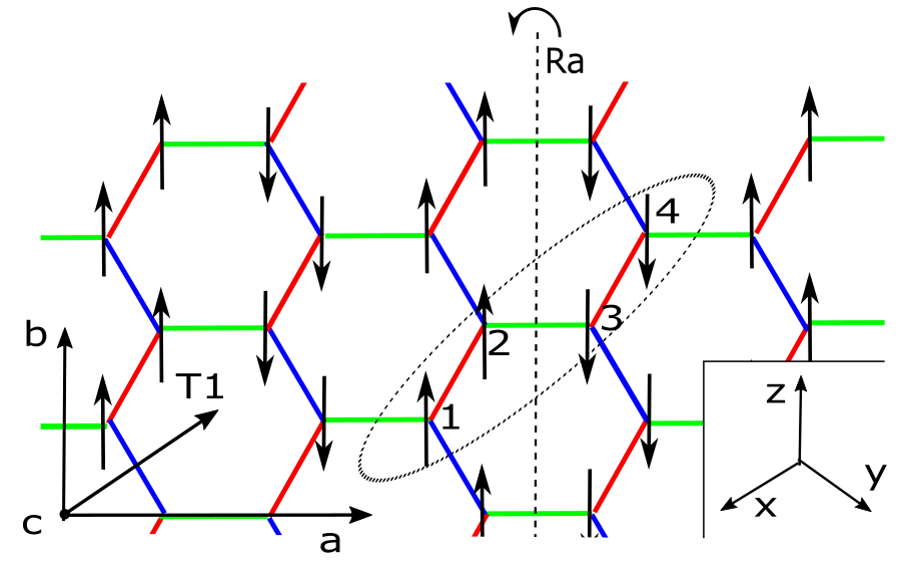}
  \caption{(Color online) Illustration of the zigzag order on the honeycomb lattice and associated symmetries. The dashed oval denotes the doubled magnetic unit cell.}
  \label{fig:zigzag order}
\end{figure}

\section{Topological magnons of the zigzag order in $\alpha$-RuCl$_3$}

While the above framework and classification applies to a LSW theory of magnons in any magnetic order, to demonstrate its power, below we apply it to one specific example: the zigzag order in layered ``Kitaev material'' $\alpha$-RuCl$_3$. In $\alpha$-RuCl$_3$ the effective spin-$1/2$'s form a quasi-2d honeycomb network, where the dominant interactions between neighboring spins are written as\cite{Rau2014,Janssen2016,Janssen2017,Winter2017b,Wu2018a}
\bea
\notag&H_{JK\Gamma h}=\sum_{\langle ij\rangle\in \alpha\beta (\gamma)} \big[KS^{\gamma}_{i}S^{\gamma}_{j}+\Gamma(S^{\alpha}_{i}S^{\beta}_{j}+S^{\beta}_{i}S^{\alpha}_{j})\big]\\
&+J\sum_{\langle ij\rangle}\vec S_i\cdot\vec S_j-\sum_i\vec h\cdot\vec S_i.\label{model:JKGammaH}
\eea
where $J$, $K$ and $\Gamma$ denote the strength of nearest-neighbor (NN) Heisenberg, Kitaev and symmetric anisotropy terms. Without external fields, a ``zigzag'' magnetic order develops as illustrated in FIG. \ref{fig:zigzag order}. Although the Bravais lattice translation $T_1$ is broken, its combination with time reversal \ie magnetic translation $\tilde T_1=T_1\cdot\bst$ is preserved by the zigzag order. Mirror reflection $R_a$ w.r.t. [100] plane is also preserved, where we have chosen the Bravais lattice vectors as $\hat a\parallel(-1,1,0),~\hat b\parallel(-1,-1,2),~\hat c\parallel(1,1,1)$.

\begin{figure*}%
\centering
\begin{subfigure}{.8\columnwidth}
\includegraphics[width=\columnwidth]{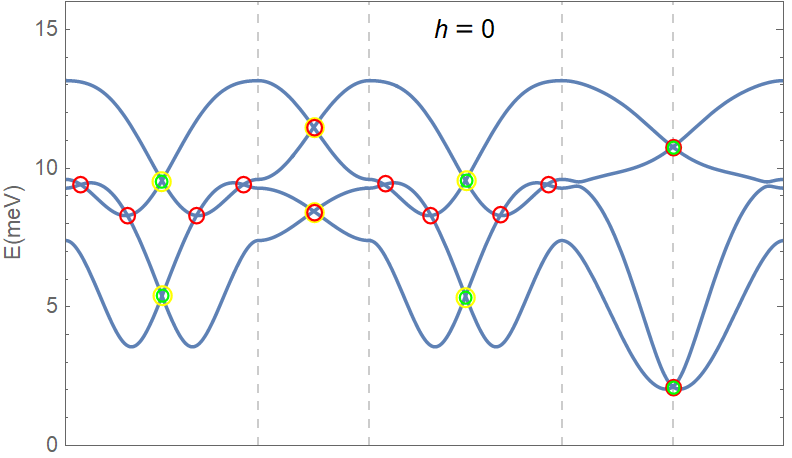}%
\caption{ }%
\label{fig:symmetrya}%
\end{subfigure}\quad\quad%
\begin{subfigure}{.8\columnwidth}
\includegraphics[width=\columnwidth]{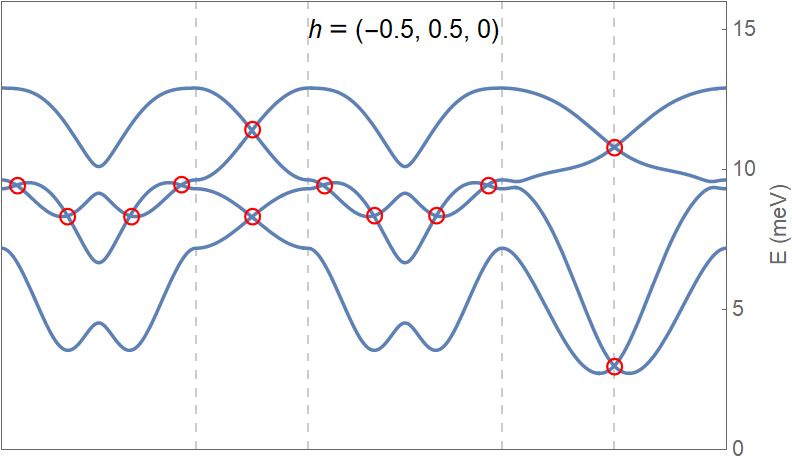}%
\caption{ }%
\label{fig:symmetryb}%
\end{subfigure}\\%
\begin{subfigure}{.8\columnwidth}
\includegraphics[width=\columnwidth]{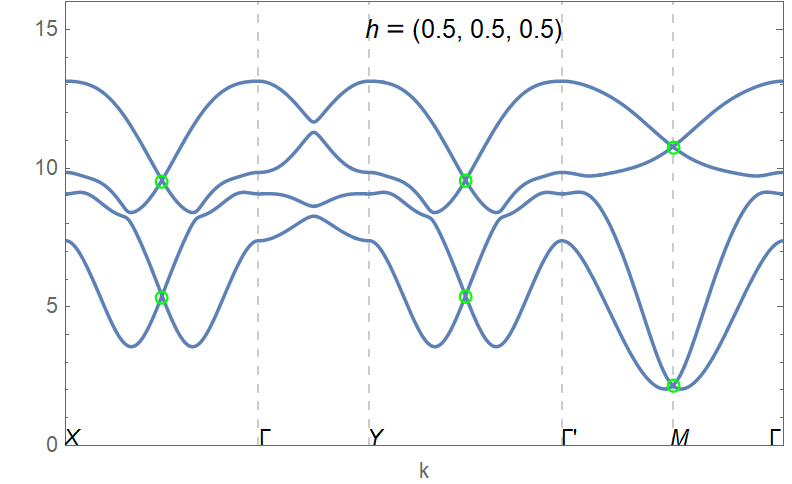}%
\caption{ }%
\label{fig:symmetryc}%
\end{subfigure}\quad\quad%
\begin{subfigure}{.8\columnwidth}
\includegraphics[width=\columnwidth]{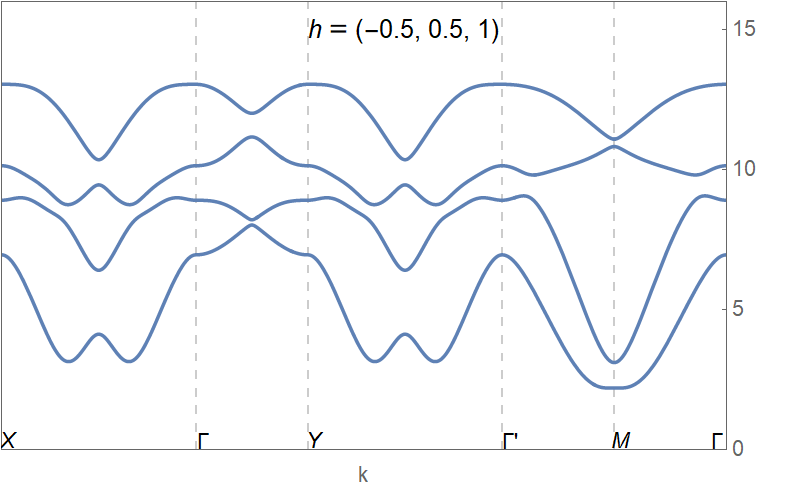}%
\caption{ }%
\label{fig:symmetryd}%
\end{subfigure}%
\caption{(Color online) Magnon band structures of zigzag-ordered $\alpha$-RuCl$_3$ under different magnetic fields, using parameters $K=-6.8$~meV, $\Gamma=9.5$~meV and $J\approx0$ in model (\ref{model:JKGammaH}) fitted from neutron data\cite{Ran2017}. Three types of magnon band crossings are protected by mirror $R_a$ (red circles), magnetic translation $\tilde T_1=\bst\cdot T_1$ (yellow circles) and magnetic mirror $\tilde R_a=R_a\cdot\tilde T_1$ (green circles).}
\label{fig:symmetry}
\end{figure*}

Using parameters $K=-6.8$~meV, $\Gamma=9.5$~meV and $J\approx0$ in model (\ref{model:JKGammaH}) from fitting recent neutron scattering data\cite{Ran2017}, we plot the magnon band structure (for details see Supplemental Materials) of zigzag-ordered $\alpha$-RuCl$_3$ in FIG. \ref{fig:symmetry}. In the absence of external fields (FIG. \ref{fig:symmetrya}), there are three types of symmetry-protected magnon band crossings, protected by mirror $R_a$ (red), magnetic translation $\tilde T_1=\bst\cdot T_1$ (yellow) and magnetic mirror $\tilde R_a\equiv R_a\cdot\tilde T_1$ (green). A magnetic field along $\hat a$-axis breaks $\tilde T_1$ but preserves mirror $R_a$, leaving only the red-colored band crossings in FIG. \ref{fig:symmetryb}. In contrast, an out-of-plane field along $\hat c$-axis breaks both $\tilde T_1$ and $R_a$ but preserves the magnetic mirror $\tilde R_a$, leaving only the green-colored band crossings in FIG. \ref{fig:symmetryc}. Finally, a generic magnetic field along a low-symmetry direction will break all symmetries and lift all the magnon band touchings, as shown in FIG. \ref{fig:symmetryd}.

\begin{figure*}%
\centering
\begin{subfigure}{.9\columnwidth}
\includegraphics[width=\columnwidth]{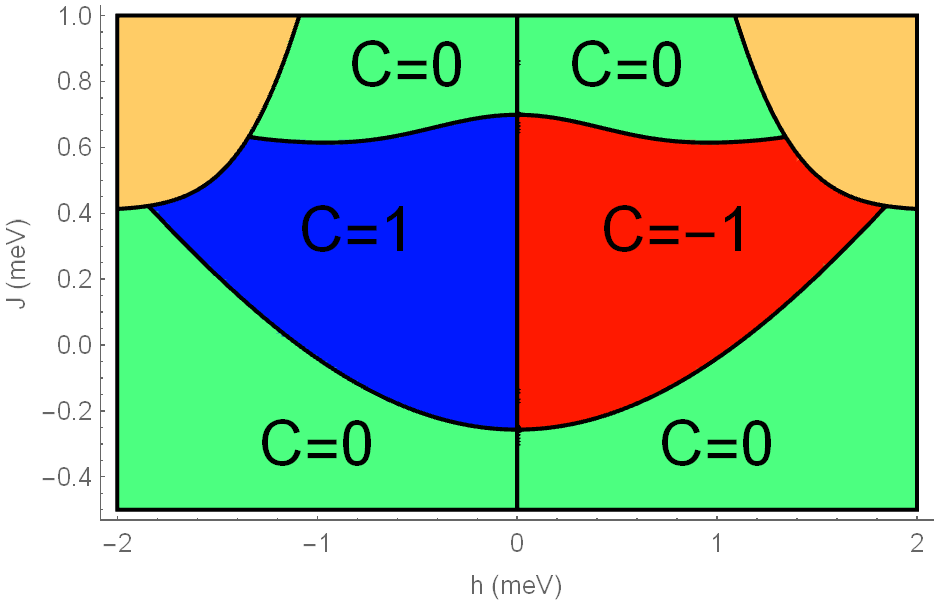}%
\caption{In-plane field along $(\hat a+\hat b)$ direction}%
\label{fig:phase1}%
\end{subfigure}\quad\quad%
\begin{subfigure}{.9\columnwidth}
\includegraphics[width=\columnwidth]{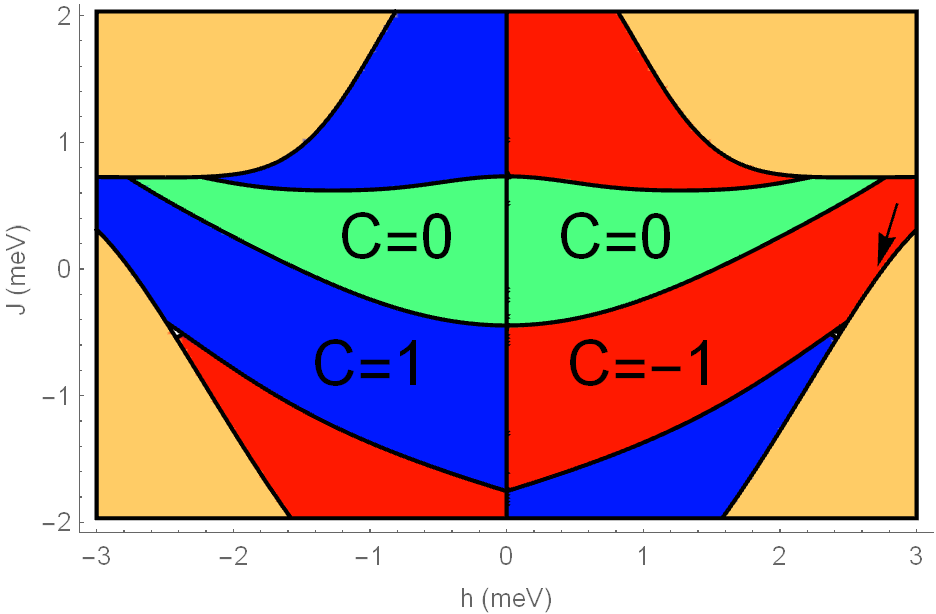}%
\caption{Out-of-plane field along (1,1,1)}%
\label{fig:phase2}%
\end{subfigure}%
\caption{(Color online) The Chern number $C$ of the lowest-energy magnon band (in FIG. \ref{fig:symmetryd}) of the zigzag order as a function of NN Heisenberg interaction $J$ and magnetic field $\vec h$ in model (\ref{model:JKGammaH}) with $K=-6.8$~meV, $\Gamma=9.5$~meV\cite{Ran2017}. The in-plane field aims at 45 degrees with both $\hat a$ and $\hat b$ axes. The zigzag order is unstable in the yellow regions.}
\label{fig:phase}
\end{figure*}

\begin{figure}%
\centering
\includegraphics[width=0.9\columnwidth]{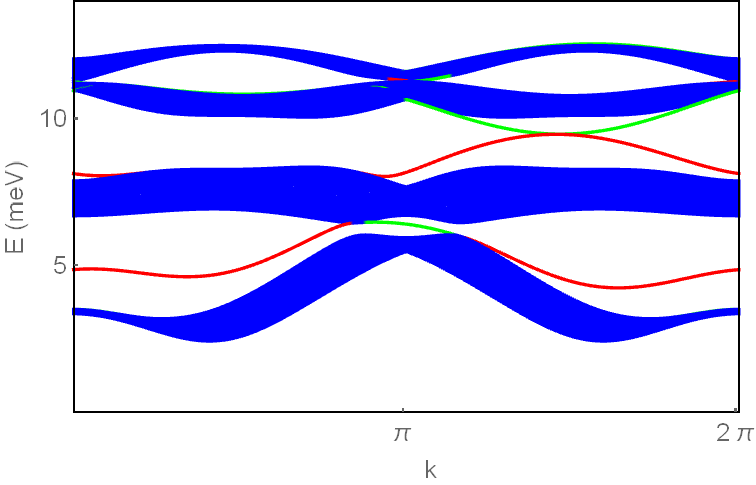}%
\caption{(Color online) Magnon spectrum on a cylinder with length $L_a=30$, periodic along $\hat b$ axis and open along $\hat a$ direction. Red and green denote topological edge states on the right and left edges respectively. We choose $J=0$ and $h_c=2.7$~meV in model (\ref{model:JKGammaH}), shown by the arrow in FIG. \ref{fig:phase2}.}
\label{fig:boundary}
\end{figure}

After the magnetic field breaks all symmetries and lifts the band crossings, the topology of each magnon band is well-defined. In the absence of symmetries, each magnon band is characterized by an integer-valued Chern number $C\in\mbz$ as shown in TABLE \ref{tab:classification}. Using the fermionization map (\ref{fermionization map}) we can numerically compute\cite{Fukui2005} the Chern number for each magnon band from the fermionized Hamiltonian ${\bf H}_f$. FIG. \ref{fig:phase} shows how the Chern number $C$ for the lowest energy magnon band (see FIG. \ref{fig:symmetryd}) depends on NN Heisenberg interaction $J$ and magnetic field $\vec h$ in model (\ref{model:JKGammaH}). Due to the bulk-boundary correspondence, $C$ is also the number of chiral magnon edge states between the lowest energy band and the one above it. Choosing $J=0$ and out-of-plane field $h_c=2.7$~meV with Chern number $C=-1$ (see the arrow in FIG. \ref{fig:phase2}), we show the magnon spectrum on a cylinder geometry in FIG. \ref{fig:boundary}, where each edge hosts a chiral magnon edge mode connecting the lowest magnon band and the one above it.

In Ref.\onlinecite{Winter2017b} the importance of anharmonic interactions between magnons beyond LSW theory has been argued for the zero-field zigzag order in $\alpha$-RuCl$_3$. Since the interactions between magnons preserve the remaining symmetries of the magnetic order, the symmetry-protected topological magnons and surface states should be stable against certain amount of anharmonicity beyond LSW theory. While most results in the main text are obtained using parameters fitted from \Ref{Ran2017}, in supplemental materials we also computed the magnon band topology for the model proposed in \Ref{Winter2017b} for comparison, where magnon Chern bands are also induced by small magnetic fields along a range of directions.

\section{Summary}

We develop a fermionization approach which maps any LSW theory to a short-ranged free-fermion Hamiltonian with exactly the same spectrum, while preserving all symmetries and band topology of the system. This allows us to classify and compute magnon band topology in various magnetic orders, hence providing a useful guide to the search for topological magnon bands and protected surface magnons in magnetic materials. Moreover this formulation can also be applied to classify and characterize the topology of various Bose-Einstein condensates.

As an application of this formulation, we investigate the zigzag magnetic order in layered honeycomb Kitaev material $\alpha$-RuCl$_3$. We identify symmetry-protected magnon band touchings at zero field, and magnon Chern bands under a small field along a wide range of directions. While recently the possibility of magnon Chern bands in $\alpha$-RuCl$_3$ has been proposed in the large field limit\cite{McClarty2018}, our studies reveal that a small field is enough to induce topological magnon bands. Our results provide a motivation for future neutron scattering and optical measurements to detect topological magnons in $\alpha$-RuCl$_3$ under a small magnetic field.

\acknowledgments We thank Pontus Laurell and Rolando Valdes Aguilar for helpful discussions. YML thanks Aspen Center for Physics for hospitality, where this draft is finalized. This work is supported by the Center for Emergent
Materials, an NSF MRSEC, under award number DMR-1420451 (FL), by NSF under award number DMR-1653769 (YML) and in part by NSF grant PHY-1607611 (YML).


%

\begin{widetext}

\newpage

\begin{center}
{\bf {Supplemental Materials}}
\end{center}

\appendix

\section{General setup of linear spin-wave theory}

\subsection{Equation-of-motion approach to spin waves}

Consider a generic bilinear Hamiltonian of a spin system $\{{\bf S}_i=(S_i^x,S_i^y,S_i^z)|i\in L\}$ on lattice $L$ ($\hbar$ is set to unity unless specifically mentioned):
\bea\label{mag ham}
&H_0=\frac12\sum_{i,j\in L}\sum_{\alpha,\beta} S_i^\alpha g_{i,j}^{\alpha,\beta} S_j^\beta,\\
\notag&g_{i,j}^{\alpha,\beta}=g_{j,i}^{\beta,\alpha},~~|{\bf S}_i|=\bar S_i.
\eea
The spin magnitude $\bar S_i$ on different lattice sites $i\in L$ can in principle be different. In the classical (large spin) limit its ground state is magnetically ordered:
\bea\label{ordered moments}
\expval{{\bf S}_i}=\bar S_i\hat z_i=(0,0,\bar S_i).
\eea
where $\hat z_i$ is the unit vector along the direction of ordered moment on site $i$. We've chosen a ``local'' coordinate frame $\{\hat x_i,\hat y_i\perp\hat x_i,\hat z_i=\hat x_i\times\hat y_i\}$ according to local ordering direction, so generically our couplings $g_{i,j}^{\alpha,\beta}$ in (\ref{mag ham}) are quite different from the usual couplings in a global Cartesian coordinate frame where $\hat x_i=\hat x_j,~\forall~i,j$. For simplicity we choose all these local frames to be right-handed just like in the global frame. For example in our notation, an isotropic Heisenberg model with magnetic coupling $J_{i,j}$ is given by
\bea
g^{\alpha,\beta}_{i,j}=J_{i,j}\hat\alpha_i\cdot\hat\beta_j,~~~\alpha,\beta=x,y,z.
\eea
Without loss of generality, a stable classical ground state (\ref{ordered moments}) must minimize the classical energy (\ref{mag ham}), under the constraints \bea
{\bf S}_i\cdot{\bf S}_i=(\bar S_i)^2,~~~\forall~i\in L.
\eea
which can be enforced by introducing a Lagrangian multiplier per site $\{\lambda_i|i\in L\}$. In other words we need to solve the minimization problem of ``constrained Hamiltonian'' $F=H_0+$constraints. To be specific, a stable magnetic order (\ref{ordered moments}) must satisfy the saddle-point condition
\bea\label{saddle point eq.}
&\frac{\partial F}{\partial S_i^\alpha}|_{{\bf S}_i=\expval{{\bf S}_i}}=\sum_{j,\beta}g_{i,j}^{\alpha,\beta} \expval{S_j^\beta}+2\lambda_i\expval{S_i^\alpha}=0,\\
&F[{\bf S}_i,\lambda_i]=\frac12\sum_{i,j}\sum_{\alpha,\beta} S_i^\alpha g_{i,j}^{\alpha,\beta} S_j^\beta+\sum_i\lambda_i(|{\bf S}_i|^2-\bar S_i^2).\notag
\eea
and the stability condition which guarantees positive stiffness for the order
\bea
\notag&\sum_{i,j}\sum_{\alpha,\beta}\delta S_i^\alpha\frac{\partial^2 F}{\partial S_i^\alpha\partial S_j^\beta}|_{{\bf S}_i=\expval{{\bf S}_i}}\delta S_j^\beta=\\
&\delta S_i^\alpha g_{i,j}^{\alpha,\beta}\delta S_j^\beta+2\sum_i\lambda_i\sum_\alpha(\delta S_i^\alpha)^2\geq0,~~~\forall~\{\delta{\bf S}_i\}.\label{minimization condition}
\eea
The values of Lagrangian multipliers $\{\lambda_i\}$ are determined by requiring
\bea
\frac{\partial F}{\partial\lambda_i}|_{{\bf S}_i=\expval{{\bf S}_i}}=(\expval{{\bf S}_i})^2-(\bar S_i)^2=0.
\eea\\

The low-energy dynamics of the ordered magnets is captured by the spin waves, \ie small deviations of spins from their ordered moments (\ref{ordered moments}):
\bea\label{spin wave variables}
{\bf s}_i\equiv{\bf S}_i-\expval{{\bf S}_i}\Rightarrow\expval{{\bf S}_i}\perp{\bf s}_i=(s_i^x,s_i^y,0),~~|{\bf s}_i|\ll\bar S_i.
\eea
Using the commutation relations for spin operators
\bea
[S_i^\alpha,S_j^\beta]=\imth\hbar\epsilon_{\alpha\beta\gamma}S_i^\gamma~\delta_{i,j}
\eea
we can obtain their linearized equations of motion (repeated Greek indices are summed over)
\bea
&\notag\frac{\text{d}s_i^\alpha}{\text{d}t}=\frac{[S_i^\alpha,H_0]}{\imth\hbar}=\sum_{j}\epsilon_{\alpha\mu\nu}S_i^\nu g^{\mu,\beta}_{i,j}S_j^\beta\\
&\notag=\sum_j\epsilon_{\alpha\mu\nu}g^{\mu,\beta}_{i,j}(\expval{S_i^\nu}s_j^\beta+s_i^\nu\expval{S_j^\beta})\\
&\notag=\epsilon_{\alpha\mu\nu}\expval{S_i^\nu}(2\lambda_is_i^\mu+\sum_jg^{\mu,\beta}_{i,j}s_j^\beta)\\
&\notag=\bar S_i\epsilon_{\alpha\mu z}(2\lambda_is_i^\mu+\sum_jg^{\mu,\beta}_{i,j}s_j^\beta)=\sum_{j,\beta}M_{i\alpha,j\beta}s_j^\beta,\\
\notag&{\bf M}={\bf A}\cdot{\bf R},~~~A_{i\alpha,j\beta}=\epsilon_{\alpha\beta}\delta_{i,j}\bar S_i=(\imth\sigma_y)_{\alpha\beta}\delta_{i,j}\bar S_i,\\
&R_{i\alpha,j\beta}=2\lambda_{i}\delta_{i,j}\delta_{\alpha,\beta}+g_{i,j}^{\alpha,\beta}\label{spin wave eom:raw}
\eea
where we have used (\ref{ordered moments}) and saddle-point condition (\ref{saddle point eq.}), and $\vec\sigma$ are Pauli matrices in the $(s^x,s^y)$ space. Clearly ${\bf A}$ is a $2N\times2N$ ($N$ being the total number of spins) skew-symmetric real matrix, while ${\bf R}$ is a $2N\times2N$ real symmetric matrix. According to stability condition (\ref{minimization condition}), $R$ must be nonnegative definite \ie its eigenvalues must be either zero or positive. By rescaling the spin wave variables to be ${\bf s}_i\rightarrow{\bf s}_i/\sqrt{\bar S_i}$ we can always rewrite the spin wave equations (\ref{spin wave eom:raw}) into the standard form
\bea
&\notag {\bf A}=\imth\sigma_y,\\
& {\bf R}_{{i\alpha,j\beta}}=\sqrt{\bar S_i}(g_{i,j}^{\alpha,\beta}+2\lambda_{i}\delta_{i,j}\delta_{\alpha,\beta})\sqrt{\bar S_j}.\label{standard form}
\eea
where ${\bf R}$ remains a real symmetric nonnegative-definite matrix, describing short-ranged spin-spin interactions in the physical system. From now on we will refer ${\bf R}$ as the spin-wave (or magnon) Hamiltonian.

It is straightforward to show the above analysis applies even beyond quadratic magnetic Hamiltonian (\ref{mag ham}). For example in the presence of Zeeman field $\{h_i^\alpha\}$, the same analysis leads to the following real symmetric magnon Hamiltonian:
\bea
{\bf R}_{{i\alpha,j\beta}}=\sqrt{\bar S_i}\big[g_{i,j}^{\alpha,\beta}+(2\lambda_{i}\delta_{\alpha,\beta}-h_i^\alpha)\delta_{i,j}\big]\sqrt{\bar S_j}
\eea
which is stable only if ${\bf R}$ is non-negative definite.

As a result, quite generally, the following equation of motion (e.o.m.) determines the dynamics of spin waves
\bea\label{spin wave eom}
-\imth\frac{\text{d}s_i^\alpha}{\text{d}t}=\sum_{j,\beta}\big(\sigma_y\cdot{\bf R}\big)_{i\alpha,j\beta}s_j^\beta
\eea
In this e.o.m. approach, eigenvalues of matrix $\sigma_y\cdot{\bf R}$ determine the magnon (or spin wave) spectrum. The fact that ${\bf R}={\bf R}^\ast$ is a real matrix implies a ``particle-hole symmetry'' of the eigenvalues of $\sigma_y{\bf R}$: a positive eigenvalue $\omega_j>0$ must appear in pair with a negative eigenvalue $-\omega_j<0$.

\subsection{Holstein-Primakoff approach to spin waves}

In Holstein-Primakoff formulation\cite{Holstein1940}, the spin wave dynamics is described in a boson representation of ordered spins in the semiclassical large-$S$ limit. Here we show that the Holstein-Primakoff approach to spin waves is in fact equivalent to the e.o.m. approach introduced earlier, and the two approaches are simply related by a unitary rotation $U$ in (\ref{unitary rot}).

In the local frame (\ref{ordered moments}) of ordered magnetic moments, the spin operators are written in terms of boson annihilation and creation operators $\{b_i,b_i^\dagger\}$
\bea
&\notag S^z_i=\bar S_i-b_i^\dagger b_i,\\
&S^+_i\equiv S_i^x+\imth S_i^y=\sqrt{2\bar S_i-b_i^\dagger b_i}\cdot b_i.
\eea
It's straightforward to verify the spin commutation relation under the condition
\bea
b^\dagger_ib_i\leq2\bar S_i.
\eea
A large-$S$ expansion in ${b_i^\dagger b_i}/{\bar S_i}\ll1$ leads to
\bea
\notag&\bpm\hat s_i^x\\ \hat s_i^y\epm=\sqrt{{\bar S_i}}\cdot U\bpm b_i\\b_i^\dagger\epm+O(b^3),\\
&U=\frac1{\sqrt2}\bpm1&1\\-\imth&\imth\epm=e^{\imth\frac\pi4(1-\sigma_z)}e^{\imth\frac\pi4\sigma_y}.\label{unitary rot}
\eea
Now we can perform large-$S$ expansion on Hamiltonian (\ref{mag ham}) and only keep terms up to quadratic order in $b,b^\dagger$. The result is the following linearized equation
\bea
&\notag\hat H_{eff}=\frac12\sum_{i,j}\sum_{\alpha,\beta=x,y}\hat s_i^\alpha g_{i,j}^{\alpha,\beta}\hat s_j^\beta+2\sum_i\lambda_ib_i^\dagger b_i\\
&=\frac12\sum_{i,j}\sum_{\alpha,\beta}\hat s_i^\alpha R_{i\alpha,j\beta}\hat s_j^\beta+\text{const.}\label{h-p ham}
\eea
where we have used the saddle-point condition (\ref{saddle point eq.}) \ie
\bea
\sum_{j}g_{i,j}^{\alpha,z} \expval{S_j^z}=-2\lambda_i\expval{S_i^z}\delta_{\alpha,z}
\eea
in our local coordinate frame. In terms of bosons the spin wave Hamiltonian (\ref{h-p ham}) is written as
\bea
&\notag\hat H_\text{H-P}=\sum_{i,j}\phi^\dagger_i({\bf H}_{\bf b})_{i,j}\phi_j,\\
&\notag\phi_j=(\phi_{j,1},\phi_{j,2})^T\equiv(b_j,b_j^\dagger)^T,\\
&({\bf H}_{\bf b})_{i,j}=U^\dagger\sqrt{\bar S_i}R_{i,j}\sqrt{\bar S_j}U=U^\dagger{\bf R}_{i,j}U\label{h-p ham:relation to spin wave}
\eea

The fact that ${\bf R}$ is a real symmetric matrix imposes the following constraint on non-negative-definite Hamiltonian ${\bf H}$:
\bea\label{condition:real}
({\bf H}_b)^\ast=\sigma_x{\bf H}_b\sigma_x
\eea
since $U^\ast=U\sigma_x=\sigma_zU$.

\subsection{Structure of the magnon spectrum}

Holstein-Primakoff Hamiltonian (\ref{h-p ham:relation to spin wave}) is generally a Bogoliubov-de Gennes (BdG) Hamiltonian of boson operators $\{b_i,b_i^\dagger\}$, involving both quadratic hopping and pairing terms of bosons. To diagonalize the boson BdG Hamiltonian ${\bf H}_b$ in (\ref{h-p ham:relation to spin wave}), one needs to find a Bogoliubov transformation ${\bf W}_b\in SU(N,N)$ such that
\bea
\notag&{\bf W}_b^\dagger{\bf H}_b{\bf W}_b={\bf \Lambda},~~~{\bf \Lambda}_{i\alpha,j\beta}=\lambda_{i\alpha}\delta_{i,j}\delta_{\alpha,\beta},\\
&{\bf W}_b^\dagger\sigma_z{\bf W}_b=\sigma_z\label{condition:simplectic}.
\eea
where the 2nd condition ${\bf W}_b^\dagger\sigma_z{\bf W}_b=\sigma_z$ guarantees the boson commutation relation
\bea
[b_i,b^\dagger_j]=\delta_{i,j}\Leftrightarrow[\phi_{i\alpha},\phi_{j\beta}^\dagger]=(\sigma_z)_{\alpha,\beta}\delta_{i,j}.
\eea
remains invariant under the Bogoliubov transformation ${\bf W}_b$. It's straightforward to show that we are effectively digonalizing matrix $\sigma_z{\bf H}_b$ since
\bea
\sigma_z{\bf H}_b\cdot{\bf W}_b={\bf W}_b\cdot\sigma_z{\bf\Lambda}
\eea
Condition (\ref{condition:real}) guarantees that eigenstates with opposite frequency $\pm E$ always show up in pairs:
\bea
\sigma_z{\bf H}_b\cdot\vec v_E=E~\vec v_E\Longleftrightarrow\sigma_z{\bf H}_b\cdot\sigma_x\vec v_E^\ast=-E~\sigma_x\vec v_E^\ast.\notag
\eea
in other words
\bea
{\bf W}_b=({\bf V},\sigma_x{\bf V}^\ast)=\sigma_x{\bf W}_b^\ast\sigma_x.
\eea
where ${\bf V}$ is a $2N\times N$ matrix satisfying normalization condition
\bea
{\bf V}^\dagger\sigma_z{\bf V}=\hat1_{N\times N}.
\eea
$N$ being the total number of spins. This means each eigenvalue in ${\bf \Lambda}$ is at least two-fold degenerate \ie
\bea
{\bf\Lambda}=\sigma_0\otimes{\bf\Omega},~~~\Omega_{i,j}=\delta_{i,j}\omega_i\geq0.
\eea
Therefore condition (\ref{condition:real}), equivalent to condition ${\bf R}={\bf R}^\ast$ in  can be viewed as a ``particle-hole symmetry'' which relates positive-eigenvalue ($+\omega_i$) states to negative-eigenvalue ($-\omega_i$) ones.

Clearly matrix $\sigma_z{\bf H}_b$ has the same magnon spectrum as the equation of motion (e.o.m.) approach (\ref{spin wave eom}) to spin waves, since
\bea
&\notag\sigma_z{\bf H}_b=\sigma_zU^\dagger{\bf R}U\\
&=U^{-1}\big(U\sigma_zU^\dagger{\bf R}\big)U=U^{-1}(\sigma_y{\bf R})U
\eea
The corresponding basis transformation ${\bf B}$ that diagonalizes spin wave equation of motion (\ref{spin wave eom}) is
\bea
{\bf B}=U{\bf W}U^\dagger\Longrightarrow\sigma_y{\bf R}\cdot{\bf B}={\bf B}\cdot(\sigma_y\otimes{\bf\Omega})
\eea
It's straightforward to check the following properties for ${\bf B}$ from (\ref{condition:simplectic}) and (\ref{condition:real}):
\bea
{\bf B}^\dagger\sigma_y{\bf B}=\sigma_y,~~~{\bf B}^\ast={\bf B}.
\eea
Therefore ${\bf B}$ is a real symplectic matrix, which diagonalizes the non-negative-definite matrix ${\bf R}$ in (\ref{spin wave eom}) by:
\bea\label{diagonal form of h-p ham}
&{\bf B}^\dagger{\bf R}{\bf B}={\bf\Lambda}=\sigma_0\otimes{\bf\Omega},~~~{\bf B}\in Sp(2N,\mathbb{R}),\\
&\Leftrightarrow\sigma_y{\bf R}={\bf B}(\sigma_y\otimes{\bf\Omega}){\bf B}^{-1}={\bf B}{\bf\Omega}{\bf B}^\dagger\sigma_y.\notag
\eea
The equivalence between e.o.m. approach and Holstein-Primakoff approach to spin waves is summarized in TABLE \ref{tab:map}.

\section{Implementing symmetries in spin waves}

The many-spin Hamiltonian (\ref{mag ham}) can preserve various symmetries, such as global $SO(3)$ spin rotations, time reversal $\bst$ and space group symmetries. We call this symmetry group $G_0$. Formation of magnetic orders generally breaks the original symmetry $G_0$ down to a subgroup $G_s\subset G_0$, which does not include global time reversal symmetry $\bst$. However the combination of time reversal and another operation may still be a symmetry even in the presence of the magnetic order: \eg a collinear ferromagnetic order preserves the combination of time reversal $\bst$, and a spin rotation $e^{\imth\pi\sum_j\hat S_j^x}$ by angle $\pi$ along an axis perpendicular to the ordering direction. An antiferromagnetic Neel order on a bipartite lattice typically preserves the combination of $\bst$, and some space group operation exchanging two sublattices, such as translation on a square lattice.

In the following we discuss how these unbroken symmetries act on a spin-wave system (\ref{spin wave eom}) or (\ref{h-p ham:relation to spin wave}). A generic symmetry $g$ is implemented on a magnon Hamiltonian ${\bf R}$ in the following way:
\bea\label{sym:generic}
{\bf R}\overset{g}\longrightarrow O_g{\bf R}O_g^\dagger,~~~O_g\in SO(2N),~~~\forall~g\in G_s.
\eea

A symmetry element $g\in G_s$ can be either unitary such as spin rotations and crystalline symmetries, or anti-unitary such as time reversal or its combination with a unitary operation. They have quite different effects on the spin waves. In particular while a unitary symmetry preserves the handedness of the local coordinate frame for the ordering moments, an anti-unitary symmetry switches the handedness, since time reversal operation $\bst$ will reverse all spin components ${\bf S}_i\overset{\bst}\longrightarrow-{\bf S}_i$. In terms of their matrix representation $O_g$ in (\ref{sym:generic}), they differ in the following way:
\bea
[O_g,\sigma_y]=0,~~~\forall~g~\text{is unitary}\Leftrightarrow O_g\in SO(2N)\cap Sp(2N,\mathbb{R})\simeq U(N).
\eea
and
\bea
\{O_g,\sigma_y\}=0,~~~\forall~g~\text{is anti-unitary}.
\eea
It's straightforward to see that the spin-wave e.o.m. (\ref{spin wave eom}) remains invariant under either a unitary or anti-unitary symmetry $g$, since time reverses $t\overset{\bst}\rightarrow-t$ under an anti-unitary symmetry $g$.

As show in TABLE \ref{tab:map}, the symmetry operations $\{O_g|g\in G_s\}$ on a ``magnon Hamiltonian'' ${\bf R}$ can also be translated into the associated Holstein-Primakoff formalism, and the corresponding free fermion system. In particular, each unitary symmetry $g$ in a magnon system is mapped to a unitary symmetry in the fermion system; while each anti-unitary symmetry $h$ in a magnon system is mapped to an anti-unitary one in free fermions.

More specifically let's consider $h=\mathcal{K}$ (complex conjugation) and we have
\bea
&\notag\bst\imth\bst^{-1}=-\imth,~~\bst\sigma_y\bst=-\sigma_y,\\
\label{sym:time reversal}&\bst{\bf R}\bst^{-1}={\bf R},~~\bst{\bf B}\bst^{-1}={\bf B},\\
&\notag\bst U\bst^{-1}=U^\ast=U\sigma_x,~~\bst{\bf H}\bst^{-1}={\bf H}^\ast=\sigma_x{\bf H}\sigma_x.
\eea
Therefore in coplanar (and collinear) magnetic orders whose magnetic moments are all perpendicular to \eg $\hat x$-axis, the combined $\pi$-spin-rotation and time reversal symmetry is implemented by
\bea
\label{sym:coplanar}e^{\imth\pi\sum_j\hat S^x_j}\bst=\sigma_z\cdot\mathcal{K}
\eea
As a result, spin wave Hamiltonians (\ref{h-p ham}) and (\ref{h-p ham:relation to spin wave}) for coplanar magnetic orders satisfy
\bea
&\notag\sigma_z{\bf R}\sigma_z={\bf R}\Leftrightarrow\sigma_x{\bf H}\sigma_x={\bf H},\\
&\Longrightarrow\sigma_z{\bf B}\sigma_z={\bf B}.\label{condition:coplanar}
\eea
since $U^\dagger\sigma_zU=-\sigma_x$.

Spin rotational symmetries also exist in certain magnets, \eg collinear magnetic orders preserve a $U(1)$ spin rotation along the magnetization direction ($\hat z$ axis). A global spin rotation by angle $\phi$ along $\hat z$-axis is implemented by
\bea
\label{sym:collinear U(1) Sz}\hat U_z(\phi)\equiv e^{\imth\phi\sum_j\hat S_j^z}=e^{\imth\frac\phi2\sigma_y}
\eea
Therefore in collinear magnetic orders, $U(1)_{{\bf S}^z}$ spin rotational symmetry indicates
\bea
&\notag\sigma_y{\bf R}\sigma_y={\bf R}\Leftrightarrow\sigma_z{\bf H}\sigma_z={\bf H},\\
&\Longrightarrow\sigma_y{\bf B}\sigma_y={\bf B}.\label{condition:collinear U(1) Sz}
\eea

\section{Relation to previous formulation\cite{Shindou2013}}

Previously in \Ref{Shindou2013}, the Chern number of a magnon band has been computed using the projection operator into one magnon band. In particular, \Ref{Shindou2013} adopted a Cholesky decompostion\cite{Colpa1978} of the Holstein-Primakoff ${\bf H}_b$ to obtain the eigenstate wavefunctions of the LSW theory. Below we discuss the difference between our fermionization map (\ref{fermionization map}) in TABLE \ref{tab:map} and the formulation adopted in \Ref{Shindou2013}.

In \Ref{Shindou2013}, a Cholesky decomposition for Hermitian positive definite matrix ${\bf H}_b=U{\bf R}U^\dagger$ is performed
\bea
{\bf H}_b={\bf K}^\dagger\cdot{\bf K}
\eea
and leads to a free-fermion Hamiltonian
\bea
\tilde{\bf H}_f={\bf K}\cdot\sigma_z\cdot{\bf K}^\dagger
\eea
which can be diagonalized by a unitary matrix
\bea
\tilde {\bf W}_f={\bf K}{\bf W}_b{\bf \Omega}^{-1/2}
\eea
Compared to the unitary matrix ${\bf W}_f$ which diagonalizes the fermionized Hamiltonian ${\bf H}_f$ in TABLE \ref{tab:map}, this unitary transformation differs by a unitary transformation ${\bf U}_f$:
\bea
{\bf W}_f=U\sqrt{{\bf H}_b}{\bf W}_b{\bf \Omega}^{-1/2}={\bf U}_f\tilde{\bf W}_f,~~~{\bf U}_f\equiv U\sqrt{{\bf H}_b}\cdot{\bf K}^{-1},~~~U=e^{\imth\frac\pi4(1-\sigma_z)}e^{\imth\frac\pi4\sigma_y}.
\eea
It's straightforward to show that ${\bf U}_f$ is unitary
\bea
{\bf U}_f^\dagger{\bf U}_f=\big({\bf K}^\dagger\big)^{-1}{\bf H}_b{\bf K}^{-1}=1.
\eea

Though only differing by a unitary transformation, there is one major advantage of our fermionized Hamiltonian ${\bf H}_f$ over $\tilde {\bf H}_f$ previously used in \Ref{Shindou2013}. In our fermionization map, the LSW Hamiltonian matrix ${\bf R}$ and its free-fermion counterpart ${\bf H}_f$ share the same symmetry implementation $O_g$ for symmetry element $g$, where $O_g$ is independent of the Hamiltonian as long the symmetry is preserved. In contrast, for the free-fermion Hamiltonian $\tilde{\bf H}_f$ introduced in \Ref{Shindou2013}, the corresponding symmetry operation
\bea
\tilde O_g={\bf U}_f^\dagger O_g{\bf U}_f,~~~[\tilde O_g,\tilde{\bf H}_f]=0.
\eea
depends on the specific Hamiltonian ${\bf H}_b$ (or ${\bf R}$). In particular for the periodic band structure of a LSW, calculations are performed in momentum (${\bf k}$) space after the Fourier transform. Since the Holstein-Primakoff Hamiltonian ${\bf H}_b({\bf k})$ depends on momentum ${\bf k}$, the associated symmetry operator $\tilde O_g({\bf k})$ will also change with ${\bf k}$ and may not even be a smooth function of ${\bf k}$.

\section{Topological triviality of magnon ground states and Goldstone modes}

Previously we have established a mapping from a non-interacting magnon system to a free fermion system. While this map allows us to understand the band topology of magnons by examining its free-fermion counterparts, it is not a surjective map. In other words, not all free-fermion states have their counterparts in the magnon system. In this section, we establish a most significant difference between magnon and free-fermion systems. In gapped topological insulators and superconductors of fermions, the ground state can have a nontrivial topology, and host in-gap surface states between the empty conduction bands (positive) and filled valence bands (negative energy). In sharp contrast in a generic gapped magnon system, the magnetic ground state must have a trivial topology, and hence there can be no symmetry-protected in-gap surface states in any gapped magnon system. In this section, we always refer to the gap around zero energy unless further specified.

First of all, the ``wavefunction'' matrix ${\bf B}$ of spin wave Hamiltonian ${\bf R}$ has the following properties
\bea
{\bf B}^\dagger\sigma_y{\bf B}=\sigma_y,~~~{\bf B}^\ast={\bf B}
\eea
and hence belongs to the symplectic group ${\bf B}\in Sp(2N,\mathbb{R})$. As a general property of a real symplectic matrix, we have
\bea\label{symplectic:decomp}
&{\bf B}=O{\bf Z}O^\prime,~~~{\bf Z}=\bpm D&\\&D^{-1}\epm,\\
&O,O^\prime\in Sp(2N,\mathbb{R})\cap SO(2N)\cong U(N).\notag
\eea
where $D$ is a positive-definite and diagonal $N\times N$ matrix. As shown in TABLE \ref{tab:map} the spin wave ``Hamiltonian'' ${\bf R}$, a real symmetric non-negative-definite matrix, can be written as
\bea\label{diag form: spin wave}
{\bf R}=\sigma_y{\bf B}\mathbf{\Omega}{\bf B}^\dagger\sigma_y.
\eea
where $\mathbf{\Omega}$ is a non-negative-definite diagonal matrix of magnon frequencies.

Now let's assume a gapped magnon spectrum (without massless Goldstone modes) with a finite gap $2\Delta>0$ \ie
\bea
\mathbf{\Omega}_{i,j}=\delta_{i,j}\omega_i\geq0,~~~\omega_i\geq\Delta>0.
\eea

In order to study the in-gap surface states of a magnon system without loss of generality, we follow the spectrum flattening trick used in free-fermion systems. More specifically, all magnon frequencies $\{\omega_i\geq\Delta>0\}$ in the gapped spectrum are adiabatically tuned to the same positive frequency
\bea\label{flat spectrum}
\omega_i\equiv\Delta>0,~\forall~i\Longleftrightarrow\mathbf{\Omega}=\Delta\cdot\hat 1_{N\times N}.
\eea
in the ``flat band'' limit. Any zero-energy topological surface states below the gap should not be affected in this spectral flattening process.

With the flattened spectrum (\ref{flat spectrum}), due to property (\ref{symplectic:decomp}) of symplectic matrix ${\bf B}$, the spin wave Hamiltonian (\ref{diag form: spin wave}) can be written as
\bea\label{exp map}
&{\bf R}=\Delta\cdot\sigma_yO{\bf Z}^2O^\dagger\sigma_y=\Delta\cdot O{\bf Z}^{-2}O^\dagger=\Delta\cdot e^{{\bf M}_R},\\
&\notag{\bf M}_R\equiv\sigma_yO\log({\bf Z}^2)O^\dagger\sigma_y=-2O\big(\sigma_z\otimes\log D\big)O^\dagger.
\eea

The existence of an exponential map from invertible real symmetric matrix ${\bf M}_R$ to spin wave Hamiltonian ${\bf R}$ is crucial to establish the trivial topology of a magnetic ground state. It provides a continuous family of gapped magnon Hamiltonians
\bea
{\bf R}(\lambda)=\Delta\cdot e^{\lambda{\bf M}_R}=\Delta\cdot\sigma_y{\bf B}(\lambda){\bf B}^\dagger(\lambda)\sigma_y,~~0\leq\lambda\leq1.
\eea
with wavefunction matrix
\bea\label{path:wf}
{\bf B}(\lambda)=O{\bf Z}^\lambda O^\prime\in Sp(2N,\mathbb{R})
\eea
Clearly the whole family of spin wave Hamiltonians all shares the flat-band spectrum (\ref{flat spectrum}). It adiabatically connects an arbitrary gapped magnon Hamiltonian ${\bf R}={\bf R}(\lambda=1)$ to a ``trivial'' Hamiltonian that is proportional to the identity matrix
\bea\label{trivial magnon ham}
&{\bf R}(\lambda=0)=\Delta\cdot\hat 1,\\
&{\bf B}(\lambda=0)=OO^\prime\in Sp(2N,\mathbb{R})\cap SO(2N)\cong U(N).\notag
\eea

In general the magnon system can preserve certain global and/or crystalline symmetries belonging to a symmetry group $G_s$. As discussed earlier, any symmetry element $g\in G_s$ is implemented on the magnon Hamiltonian ${\bf R}$ by an orthogonal rotation $O_g$
\bea
{\bf R}=O_g{\bf R}O_g^\dagger,~~~O_g\in SO(2N).
\eea
The exponential map (\ref{exp map}) of spin-wave Hamiltonian ${\bf R}$ therefore implies that
\bea
[{\bf M}_R,O_g]=0
\eea
As a result, the whole family of gapped Hamiltonian preserves the same symmetry $g$ since
\bea
{\bf R}(\lambda)=O_g{\bf R}(\lambda)O_g^\dagger,~~~0\leq\lambda\leq1.
\eea
Hence all symmetry operations of group $G_s$, which are responsible for the protected surface states, are all preserved when magnon system ${\bf R}$ is adiabatically tuned into the trivial magnon system (\ref{trivial magnon ham}) without closing the gap.

From Table \ref{tab:map} it's straightforward to show this trivial spin-wave Hamiltonian ${\bf R}(\lambda=0)$ in (\ref{trivial magnon ham}) is mapped to a topologically-trivial free-fermion Hamiltonian
\bea
{\bf H}_f(\lambda=0)=\Delta\cdot\sigma_y
\eea
which obviously has trivial band topology and no zero-energy in-gap surface states.

Therefore by establishing a gapped family of spin-wave Hamiltonians that connects an arbitrary gapped magnon system to the trivial magnon flat bands, we proved that no symmetry-protected topological surface states exist below the finite bulk gap in a generic gapped magnon system.

One natural question follows: what about gapless magnon systems with massless Goldstone modes near zero energy in their spectra? Can they support topological surface states near zero energy? The answer is again negative. This can be understood as follows. The topology of various band touchings, such as point nodes, line nodes and fermi surfaces had been classified in free fermion systems\cite{Horava2005,Wang2012,Matsuura2013,Zhao2013} by a dimensional reduction approach. In particular the classification of stable fermi surface of codimension $d_c$ coincides with the classification of gapped free-fermion ground states (hosting zero-energy surface states) in spatial dimension $d=d_c-1$. The idea is to consider a gapped and closed submanifold of the Brillouin zone that encloses the nodal points/lines or fermi surfaces, which has dimension $d=d_c-1$. Here we can adopt exactly the same strategy in a magnon system. However as shown above, all magnon ground states must be topologically trivial, without any robust zero-energy surface states of magnon systems in any spatial dimension. As a result, the topology of massless Goldstone modes near zero energy in a magnon system must also be trivial, without any protected surface states below the bulk gap around zero energy.

\section{Classifying space and topology for each magnon band}

Previously we have shown that \emph{all negative-frequency magnon bands as a whole must be topologically trivial}. This however does not imply that each magnon band itself must also be topologically trivial. In this section, based on the mapping from spin waves to free fermions in Table \ref{tab:map}, we further show that a finite energy magnon band with arbitrary (unitary or anti-unitary) symmetries can have the same topology as a free-fermion energy band (Altland-Zirnbauer class A\cite{Altland1997}) with proper \emph{unitary} symmetries. This allows us to classify the possible topological bands and topological band touchings of spin waves, with various symmetries and in all spatial dimensions.

A spin wave spectrum $\sigma_z\otimes\mathbf{\Omega}$ always has the particle-hole symmetry relating positive and negative energy eigenstates. As proven earlier, all negative-energy bands as a whole have a trivial topology, Therefore we will focus on the magnon bands at positive energy $\{\omega_j\geq0\}$. Following Kitaev's K-theory approach to classify free-fermion systems\cite{Kitaev2009,Wen2012,Morimoto2013}, without loss of generality, we again consider the following flat-band spectrum with two flat bands $\epsilon_1>\epsilon_2>0$:
\bea
\notag&\mathbf{\Omega}=\bpm\epsilon_1&&&&&\\&\cdots&&&&\\&&\epsilon_1&&&\\&&&\epsilon_2&&\\&&&&\cdots&\\&&&&&\epsilon_2\epm\\
\notag&=\big[\epsilon_1\cdot\hat 1_{M\times M}\big]\oplus\big[\epsilon_2\cdot\hat 1_{(N-M)\times(N-M)}\big]\\
&=\Delta\cdot\hat1_{N\times N}+\epsilon\cdot\hat{\mathcal{D}}_{N\times N}.\label{path:spectrum}
\eea
where we defined
\bea
&\hat{\mathcal{D}}_{N\times N}\equiv\big[+\hat 1_{M\times M}\big]\oplus\big[-\hat 1_{(N-M)\times(N-M)}\big],\\
&\Delta\equiv\frac{\epsilon_1+\epsilon_2}2>\epsilon\equiv\frac{\epsilon_1-\epsilon_2}2>0.
\eea

Although the two bands at $\epsilon_{1,2}$ as a whole are topologically trivial, each band itself can have a nontrivial topology. This can be understood as follows. We can use the same symmetric continuous path (\ref{path:wf}) to deform the eigenstate wavefunctions of linear spin waves, while keeping the spectrum the same. This leads to the following family of symmetric linear spin wave Hamiltonian matrix
\bea
{\bf R}(\lambda)=\sigma_y{\bf B}(\lambda){\bf\Omega}{\bf B}^\dagger(\lambda)\sigma_y
\eea
which interpolates the original spin wave Hamiltonian ${\bf R}={\bf R}(\lambda=1)$ and the following simplified Hamiltonian
\bea\label{path:ham}
{\bf R}(\lambda=0)={\bf B}_0{\bf\Omega}{\bf B}_0^\dagger,~~~{\bf B}_0^\dagger={\bf B}(\lambda=0)=OO^\prime\in U(N)
\eea
Due to the two-flat-band structure of spectrum (\ref{path:spectrum}), when no other symmetries are considered, the classifying space of linear spin wave Hamiltonian (\ref{path:ham}) is given by the following Grassmannian:
\bea
\frac{U(N)}{U(M)\times U(N-M)}
\eea
This leads to a classifying space $\mathcal{C}_{d}$ without other symmetries ($d$ is the spatial dimension), belonging to symmetry class A in the Altland-Zirnbauer 10-fold way\cite{Altland1997}. This exactly match the classification of the topology of each band in the corresponding free fermion Hamiltonian, ${\bf H}_f=\sqrt{\bf R}\sigma_y\sqrt{\bf R}$ obtained from linear spin wave Hamiltonian ${\bf R}$ via the fermionization map. This demonstrates that one specific band of the free fermion Hamiltonians obtained by fermionizing the linear spin wave can realize all possible band topology within the corresponding fermion symmetry class. Therefore, we can fully classify the band topology of linear spin waves by looking into their free fermion partners obtained via the fermionization map.

%
%
%
%

\section{Linear spin wave theory for the zigzag order in $\alpha$-RuCl$_3$}

Below we describe how to use LSW theory to compute the magnon band structure for the zigzag order\cite{Little2017,Wang2017a,Cookmeyer2018} in the following model for $\alpha$-RuCl$_3$:
\bea
H_{JK\Gamma h}=\sum_{\langle ij\rangle\in \alpha\beta (\gamma)} \big[KS^{\gamma}_{i}S^{\gamma}_{j}+\Gamma(S^{\alpha}_{i}S^{\beta}_{j}+S^{\beta}_{i}S^{\alpha}_{j})\big]+J\sum_{\langle ij\rangle}\vec S_i\cdot\vec S_j+\sum_i\vec h\cdot\vec S_i.\label{ham:JKGammaH}
\eea

In the first step, we find the classical spin configuration that minimizes the free energy. Luttinger-Tisza method\cite{Luttinger1946} is widely adopted, which applies to the zero magnetic field case or a small magnetic field along specific directions. Here, we start from the zigzag magnetic order and consider four spins within one doubled magnetic unit cell. We can numerically optimize energy in eight-dimensional space of variables $\{(\theta_{i},\phi_{i})\}$, where $\theta_{i}, \phi_{i}$ are the polar and azimuthal angles for the spin orientation. Considering small deviations $\bold{X}_{\bold{k}}=(S_{1,\bold{k}}^{x},S_{1,\bold{k}}^{y},S_{2,\bold{k}}^{x},S_{2,\bold{k}}^{y},S_{3,\bold{k}}^{x},S_{3,\bold{k}}^{y},S_{4,\bold{k}}^{x},S_{4,\bold{k}}^{y})$ from the ordered (lowest-energy) magnetic moment, we expand the free energy around its minimum and obtain the following LSW energy functional
\begin{equation}
H_{2}=\sum_{\bold{k}}\bold{X}_{\bold{k}}^{\dagger}\cdot \mathcal{H}_{\bold{k}}\cdot\bold{X}_{\bold{k}}
\end{equation}
where $\mathcal{H}_{\bold{k}}$ is given by
\begin{align}
\mathcal{H}_{\bold{k}}=\left(\begin{array}{c c c c c c c c}
A & 0 & B & C & 0 & 0 & E_{1} & F_{1}\\
0 & A & C & D & 0 & 0 & F'_{1} & G_{1}\\
B^{\dagger} & C^{\dagger} & A & 0 & E & F & 0 & 0 \\
C^{\dagger} & D^{\dagger} & 0 & A & F' & G & 0 & 0 \\
0 & 0 & E^{\dagger} & F'^{\dagger} & A_{1} & 0 & B_{1} & C_{1} \\
0 & 0 & F^{\dagger} & G^{\dagger} & 0 & A_{1} & C_{1} & D_{1}\\
E_{1}^{\dagger} & F_{1}'^{\dagger} & 0 & 0 & B_{1}^{\dagger} & C_{1}^{\dagger} & A_{1} & 0\\
F_{1}^{\dagger} & G_{1}^{\dagger} & 0 & 0 & C_{1}^{\dagger} & D_{1}^{\dagger} & 0 & A_{1}
\end{array}\right)
\end{align}

\begin{figure*}%
\centering
\begin{subfigure}{.45\columnwidth}
\includegraphics[width=\columnwidth]{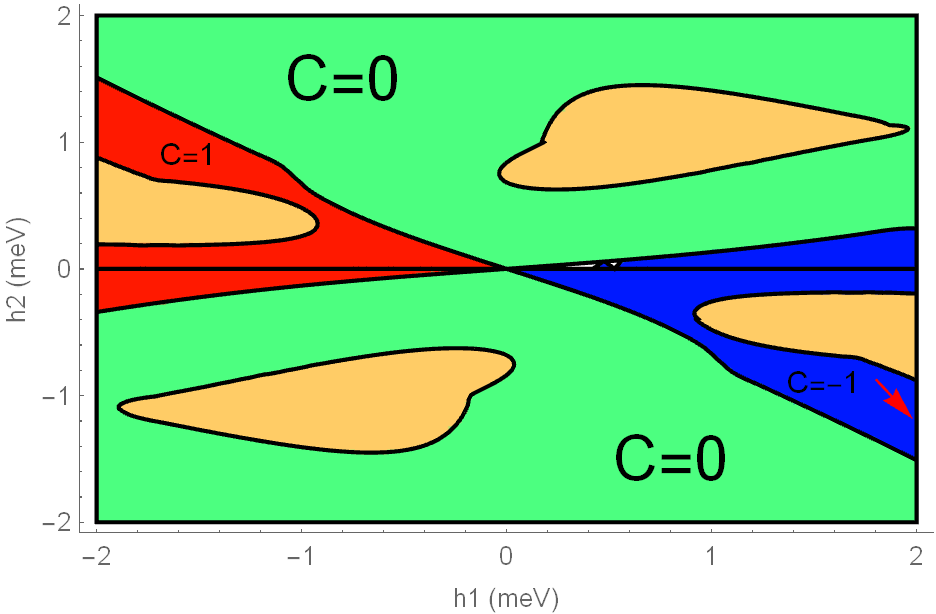}%
\caption{Field within [110] plane.}%
\label{fig:phase3}%
\end{subfigure}\quad\quad%
\begin{subfigure}{.45\columnwidth}
\includegraphics[width=\columnwidth]{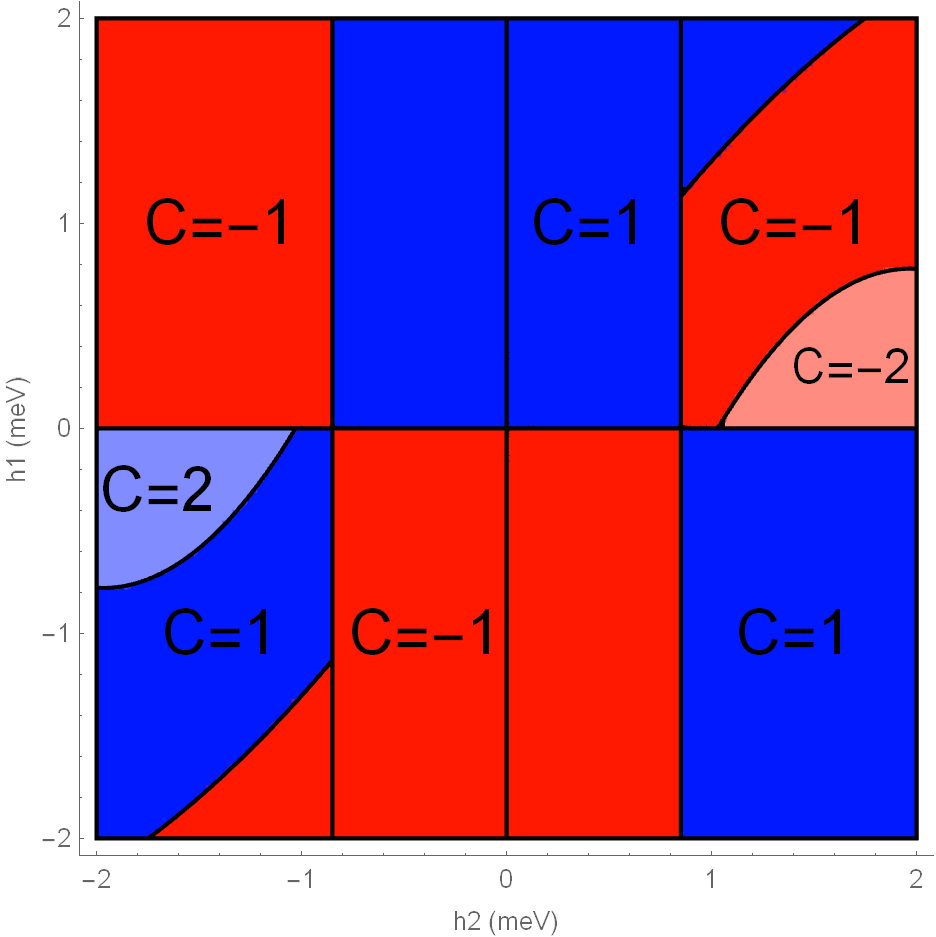}%
\caption{Field within [010] plane.}%
\label{fig:phase4}%
\end{subfigure}%
\caption{(Color online) The Chern number $C$ of the lowest-energy magnon band of the zigzag order in model (\ref{ham:JKGammaH}), with parameters $K=-5$meV, $\Gamma=2.5$meV, $J_{1}=-0.5$meV and $J_{3}=0.5$meV from \Ref{Winter2017b}. $h_1$ denotes the in-plane component and $h_2$ is the out-of-plane component of the magnetic field. The zigzag order is unstable in the yellow regions.}
\label{fig:phase3+4}
\end{figure*}

For all calculations performed in this work, we numerically found that sublattices 1 and 2 share the same ordered moment ${\bf n}$, while sublattices 3 and 4 share the opposite moment $-{\bf n}$. We represent them as $(\theta_{1},\phi_{1})$ and $(\theta_{3},\phi_{3})$. The elements of $\mathcal{H}_{\bold{k}}$ are written as below

 \begin{equation}
  \begin{aligned}
   A=&-2J-K\sin^{2}\theta_{1}+(J+K)\cos\theta_{1}\cos\theta_{3}-\Gamma \sin 2\theta_{1}(\cos \phi_{1}+\cos\phi_{3})+\sin\theta_{1}\sin\theta_{3}(J\cos(\phi_{1}-\phi_{3})\\
   &+\Gamma\sin(\phi_{1}+\phi_{3}))-2\vec{h}\cdot\hat{v}(\theta_{1},\phi_{1}),\\
   A_{1}=&-2J-K\sin^{2}\theta_{3}+(J+K)\cos\theta_{1}\cos\theta_{3}-\Gamma \sin 2\theta_{3}(\cos \phi_{1}+\cos\phi_{3})+\sin\theta_{1}\sin\theta_{3}(J\cos(\phi_{1}-\phi_{3})\\
   &+\Gamma\sin(\phi_{1}+\phi_{3}))+2\vec{h}\cdot\hat{v}(\theta_{3},\phi_{3}),\\
   B=& J(1+e^{-ik_{b}})+K(\cos^{2}\theta_{1}+e^{-ik_{b}}\sin^{2}\theta_{1}),  \\
   B_{1}=& J(1+e^{-ik_{b}})+K(\cos^{2}\theta_{3}+e^{-ik_{b}}\sin^{2}\theta_{3}),  \\
   C=& (-K\cos\theta_{1}\cos\phi_{1}-J\sin\theta_{1})\sin\phi_{1}+e^{-ik_{b}}\cos\phi_{1}(J\sin\theta_{1}+K\cos\theta_{1}\sin\phi_{1}), \\
   C_{1}=& (K\cos\theta_{3}\cos\phi_{3}+J\sin\theta_{3})\sin\phi_{3}-e^{-ik_{b}}\cos\phi_{3}(J\sin\theta_{3}+K\cos\theta_{3}\sin\phi_{3}),\\
   D=& (J+K\cos^{2}\theta_{1}\sin^{2}\phi_{1}-J\cos\phi_{1}\sin 2\theta_{1})+e^{-ik_{b}}(J+K\cos^{2}\theta_{1}\cos^{2}\phi_{1}-\Gamma\sin\phi_{1}\sin 2\theta_{1}), \\
   D_{1}=& (J+K\cos^{2}\theta_{3}\sin^{2}\phi_{3}-J\cos\phi_{3}\sin 2\theta_{3})+e^{-ik_{b}}(J+K\cos^{2}\theta_{3}\cos^{2}\phi_{3}-\Gamma\sin\phi_{3}\sin 2\theta_{3}),   \\
   E=& -J\cos(\phi_{1}-\phi_{3})+\Gamma\sin(\phi_{1}+\phi_{3}), \\
   F=& \cos\theta_{1}(-\Gamma\cos(\phi_{1}+\phi_{3})+J\sin(\phi_{1}-\phi_{3})),\\
   F'=& \cos\theta_{3}(\Gamma\cos(\phi_{1}+\phi_{3})+J\sin(\phi_{1}-\phi_{3})),\\
   G=& (J+K)\sin\theta_{1}\sin\theta_{3}+\cos\theta_{3}\cos\theta_{1}(J\cos(\phi_{1}-\phi_{3})+\Gamma\sin(\phi_{1}+\phi_{3})),\\
   E_{1}=& e^{-i(k_{a}+k_{b})}E,\quad F_{1}= e^{-i(k_{a}+k_{b})} F, \quad F_{1}'= e^{-i(k_{a}+k_{b})} F',\quad  G_{1}= e^{-i(k_{a}+k_{b})} G .
  \end{aligned}
 \end{equation}

The LSW spectrum can be obtained by diagnoalizing the above bosonic Hamiltonian.

In the main text, we use parameters $J=0$ and $K=-6.8$~meV, $\Gamma=9.5$~meV fitted from recent neutron scattering data\cite{Ran2017} in model (\ref{ham:JKGammaH}). In another recent study\cite{Winter2017b}, an ab initio guided data fit leads to a different set of parameters in model (\ref{ham:JKGammaH}): $K=-5$meV, $\Gamma=2.5$meV, $J_{1}=-0.5$meV, while including a 3rd NN Heisenberg coupling $J_{3}=0.5$meV. We have also computed magnon spectrum for this model, and obtained the Chern number $C$ of the lowest magnon band using the fermionization map. The results are summarized in FIG. \ref{fig:phase3+4}. Again a small magnetic field along a wide range of directions can give rise to a topological magnon band with a nonzero Chern number.

%
%
%
%

\end{widetext}

\end{document}